\def\g{\gamma}
\newcommand{\ul}{\underline}

\documentclass[
 reprint,
 amsmath,amssymb,
 aps, 
]{revtex4-1}

\usepackage{graphicx}
\usepackage[dvipsnames]{xcolor}
\usepackage{dcolumn}
\usepackage{bm}
\usepackage{hyperref}

\newcommand{\C}[1]{$(\ref{#1})$}

\begin{document}

\preprint{EFI-24-7}

\title{An Exact Solution for the Kinetic Ising Model with Non-Reciprocity}

\author{Gabriel Weiderpass}
\email{gaweiderpass@uchicago.edu}
\author{Mayur Sharma}%
\email{msharma3@caltech.edu}


\author{Savdeep Sethi}
\email{sethi@uchicago.edu}

 \affiliation{%
  Enrico Fermi Institute \& Kadanoff Center for Theoretical Physics \\ University of Chicago, Chicago, IL 60637, USA
}%


\date{\today}

\begin{abstract}
A wide range of non-equilibrium phenomena in nature involve non-reciprocal interactions. To understand the novel behaviors that can emerge in such systems, finding  tractable models is essential. With this goal, we introduce a non-reciprocal generalization of the kinetic Ising model in one dimension and solve it exactly. Our solution uncovers novel properties driven by non-reciprocity, such as underdamped phases, critically damped phases where a system of size 
$N$ is described by an $N^{th}$-order exceptional point, and wave phenomena influenced by the parity of $N$. Additionally, we examine the low-energy behavior of these systems in various limits, demonstrating that non-reciprocity leads to unique scaling behavior at zero temperature.
\end{abstract}

\keywords{Suggested keywords}
\maketitle

Non-reciprocal interactions are generic in systems that are out of equilibrium~\cite{aguilera_unifying_2021,fruchart_non-reciprocal_2021,ivlev_statistical_2015,you_nonreciprocity_2020,crisanti_dynamics_1987,brauns_nonreciprocal_2024,costanzo_milling-induction_2019,frohoff-hulsmann_nonreciprocal_2023,saha_scalar_2020,liu_non-reciprocal_2023,loos_irreversibility_2020,du_hidden_2024,marchetti_hydrodynamics_2013,hanai_critical_2020,hanai_non-hermitian_2019,avni2023nonreciprocal,rajeev_ising_2024,seara_non-reciprocal_2023,hanai_nonreciprocal_2024}. In a nutshell, a non-reciprocal interaction between two bodies means the response of body $A$ to body $B$ is different from the response of body $B$ to body $A$. Such interactions are seen in a large variety of systems including synthetic active matter and soft matter systems \cite{brandenbourger_non-reciprocal_2019,veenstra_non-reciprocal_2024,veenstra_non-reciprocal_2023,brandenbourger_non-reciprocal_2024,ghatak_observation_2020,helbig_generalized_2020,kotwal_active_2021,hofmann_chiral_2019,gupta_active_2022,tan_odd_2022,shankar_topological_2022,fruchart_odd_2023,colen_interpreting_2024,scheibner_odd_2020,poncet_when_2022,rosa_dynamics_2020,scheibner_non-hermitian_2020,coulais_topology_2021,dinelli_non-reciprocity_2023}, open quantum systems \cite{metelmann_nonreciprocal_2015,mcdonald_exponentially-enhanced_2020,clerk_introduction_2022,chiacchio_nonreciprocal_2023,bergholtz_exceptional_2021,begg_quantum_2024}, the collective behaviour of flocks, herds and social groups \cite{fruchart_non-reciprocal_2021,nagy_hierarchical_2010,morin_collective_2015,Ginelli-2015,hong_kuramoto_2011}, and neuroscience \cite{sompolinsky_temporal_1986,derrida_exactly_1987,parisi_asymmetric_1986,amir_non-hermitian_2016,montbrio_kuramoto_2018}. 

Given the prominent role played by non-reciprocal interactions in such a multitude of fields, including many areas of physics, finding exactly solvable many-body systems with non-reciprocity is an important endeavor of broad interest. Any such model extends the set of integrable theories studied in classical and quantum physics to this new area of high current interest. An exactly solvable model allows us to exhaustively study the effect of non-reciprocal interactions beyond numerical simulations of finite size systems. The knowledge resulting from an analytic solution can then be used as a foundation for a deeper understanding of more complex systems. 

The Ising model is the most widely studied model in statistical mechanics. It describes the thermodynamics of interacting classical spins where each spin can take two possible values. Usually one investigates the equilibrium properties of the system at fixed temperature. Studying the non-equilibrium properties of the Ising model requires a choice of dynamics for the spin degrees of freedom. There are many such choices which depend on the microscopic details of the system under investigation. 

Our starting point is the kinetic Ising model proposed by Glauber~\cite{glauber_timedependent_1963,godreche_response_2000}, which uses the principle of detailed balance to constrain the spin dynamics; see~\cite{krapivsky_kinetic_2010,henkel_non-equilibrium_2010} for nice discussions. Glauber solved this model in one spatial dimension and it has been extensively studied numerically in higher dimensions. Most discussions of the kinetic Ising model involve reciprocal interactions. However, there are also many ways to make the model non-reciprocal; some constructions studied either via mean field theory or numerical simulations appear in recent work~\cite{aguilera_unifying_2021,avni2023nonreciprocal,rajeev_ising_2024,seara_non-reciprocal_2023}.
\begin{figure}[h]
    \centering
    \includegraphics[width=.85\linewidth]{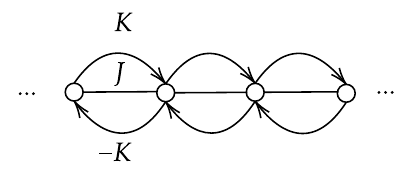}
    \caption{Ising model with non-reciprocal interactions between neighbouring spins}
    \label{NR-Ising}
\end{figure}

In this letter we begin by defining our non-reciprocal kinetic Ising model. There is no unique way to do this but our choice has the virtue of leading to a solvable model in one spatial dimension, analogous to Glauber's solution of the reciprocal kinetic Ising model. We then briefly overview how to solve the model. A more detailed discussion appears in a companion paper which also includes general results about systems where spins interact with no more than two spins, outlining the conditions under which long-time order may exist~\cite{toappear2}. 

From the exact solution, we can see a variety of new phenomena. 
For example, our non-reciprocal kinetic Ising model has a much richer phase diagram than the reciprocal case with the propagator exhibiting three distinct behaviors corresponding to overdamped, critically damped and underdamped regimes. The reciprocal kinetic Ising model corresponds to the point where the system is maximally overdamped. We also find that non-reciprocity leads to a faster relaxation to equilibrium. This is similar in spirit to observations in~\cite{ghimenti_sampling_2023}. We will also see that for many physical observables, the role of non-reciprocity is to effectively boost the system with 
\begin{align} \label{Boosts}
    x \rightarrow x + \gamma_o t \, ,
\end{align}
where $\gamma_o$ controls the strength of the non-reciprocity. 

Our non-reciprocal Ising model is schematically depicted in Figure \ref{NR-Ising}. We start with the Markov chain,
\begin{align}  \label{Markov-C}
    P\big(s(t)\big|s(t-\delta t)\big) = \prod_{j=1}^N\frac{e^{s_j(t) h_j(t-\delta t)}}{2\cosh \left(h_j(t-\delta t)\right) } \,, 
\end{align}
where $P$ is the probability of finding the spin configuration $s(t)$ given $s(t-\delta t)$. We take \C{Markov-C} as the definition of the kinetic Ising model, following \cite{aguilera_unifying_2021}. For the model of Figure \ref{NR-Ising}, the local magnetic field felt by the spin at site $j$ is
\begin{align} 
    h_j(t) = \beta J\left(s_{j-1} + s_{j+1}\right) + \beta K\left(s_{j-1} - s_{j+1}\right)  \,,
\end{align}
where $J, K$ are fixed constants and $\beta$ is the inverse temperature. The reciprocal model is $K=0$. When $K \neq 0$, it is useful to introduce the combinations $\gamma_e = \tanh\left( 2\beta J\right)$ and $\gamma_o = \tanh\left( 2 \beta K\right)$.

The basic quantities we want to determine are the equal time 
 $n$-point functions of this system,
\begin{align}
    r_{i_n ... i_1}(t) = \sum_s s_{i_n} \cdots s_{i_1} P(s,t) \,.
\end{align}
To find an equation of motion for these $n$-point functions, we use the continuous limit of \C{Markov-C} to get the master equation of this system, which we review in Appendix \ref{App-Master}. Using (\ref{1-point-eom-h}) and (\ref{tanh-in-NR}), the equation of motion for the one-point function is
\begin{align} \label{1-point-eom}
    \frac{d}{dt}r_j = - r_j + \frac{\gamma_e}{2}(r_{j-1} + r_{j+1}) + \frac{\gamma_o}{2}(r_{j-1} - r_{j+1}) \,.
\end{align}
This equation can be easily solved by Fourier analysis, 
\begin{align}
\begin{aligned} \label{1-pt-momentum}
    r_j(t) = e^{- t }&\sum_{l=1}^{ N} r_l(0) \sum_k \frac{1}{ N} \exp\Big[ik(j-l) \\
    &+ t \big( \gamma_e \cos k - i \gamma_e \sin k\big) \Big],
\end{aligned}
\end{align}
where $k=2\pi m /N$ with $m=0,1,\ldots,N-1$. This expression can be simplified using the Bessel function identity
(\ref{Sung-Generalized}),
\begin{align}
\begin{aligned}
    \label{exact-1pt}
    r_j(t) = & e^{-t} \sum_{l=1}^{ N } r_l(0) \times \\
    & \sum_{n,s=-\infty}^{\infty} (-1)^s I_{j-l+s+n N}\big( \gamma_e t\big) J_s\big(\gamma_o t\big)\,,
\end{aligned}
\end{align}
which we prove in Appendix \ref{Appendix-Bessel}. To the best of our knowledge, this identity is a new result of our work. There is an alternate method for solving \C{1-point-eom} using generating functions, which we describe in detail in \cite{toappear2}.

\begin{figure}[h]
    \centering
    \includegraphics[width=1\linewidth]{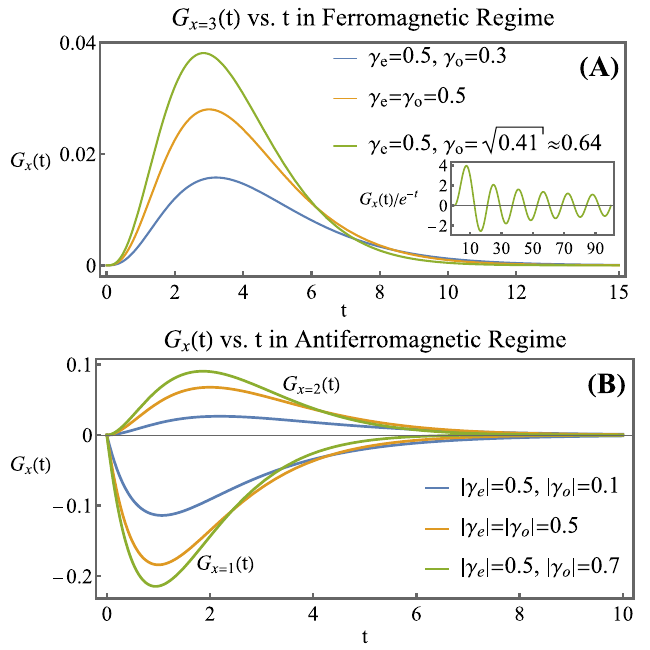}
    \caption{We plot (\ref{1-point-propagator}) as a function of time for fixed $x$. The inset in \textbf{(A)} shows that the correlation function oscillates in the underdamped regime. In the underdamped regime the correlation has a higher peak but also falls faster. In \textbf{(B)} we see the same behaviour for the antiferromagnetic case as seen in \textbf{(A)}. However when $x=\text{even}$ spins correlate while for $x=\text{odd}$, they anti-correlate.}
    \label{fig:Gx_regimes}
\end{figure}

For an infinite spin chain, we take $N\rightarrow\infty$ and drop the summation on $n$ in \C{exact-1pt}. We can also use (\ref{Graff-Neumann}) to further simplify our expression,
\begin{align}
    \label{1pt-final}
    r_j(t) & = \sum_{l=-\infty}^{\infty} r_l(0) G_{j-l}(t),
\end{align}
with the Green's function $G_x(t)$ given by
\begin{align} \label{1-point-propagator}
    G_x(t) = \left\{ \begin{aligned}
        &e^{-t} I_{x}\big( \sqrt{\gamma_e^2-\gamma_o^2} t\big) \left( \frac{\gamma_e+\gamma_o}{\gamma_e-\gamma_o}\right)^{x/2} && |\gamma_e| > |\gamma_o| \\ &c_x e^{-t} \frac{(\gamma_e t)^{|x|}}{|x|!} && |\gamma_e| = |\gamma_o| \\
        &e^{- t} J_{x}\big( \sqrt{\gamma_o^2-\gamma_e^2} t\big) \left( \frac{\gamma_e+\gamma_o}{\gamma_o-\gamma_e}\right)^{x/2} && |\gamma_e| < |\gamma_o|
    \end{aligned} \right. \,.
\end{align}
The function $c_x$ takes the form,
\begin{align}
    c_x = \left\{\begin{aligned}
       & \theta(x)  &&\gamma_e = \gamma_o\\
       & \theta(-x) &&\gamma_e=-\gamma_o
    \end{aligned}\right. \,,
\end{align}
where $c_x$ encodes the following information: if $\gamma_e=\pm\gamma_o$, excitations move to the right or left, respectively. The parameter space of the system divides into three regions,
\begin{align}
\begin{aligned}
    &|\gamma_e|>|\gamma_o|\,, \qquad \text{overdamped}\,, \\
    &|\gamma_e|=|\gamma_o|\,, \qquad \text{critically damped}\,, \\
    &|\gamma_e|<|\gamma_o|\,, \qquad \text{underdamped} \, .
\end{aligned}
\end{align}
The critically damped phase is governed by an $N^{th}$-order exceptional point of the system (\ref{1-point-eom}), which means there is one eigenvalue with degeneracy $N$ but only one associated eigenvector; see~\cite{toappear2} for a detailed discussion. Both the underdamped and critically damped phases only exist because of non-reciprocity. In Figure \ref{fig:Gx_regimes}, we illustrate these different regimes for the Green's function. 

To study finite chains with periodic or open boundary conditions we replace $G_x(t)$ in (\ref{1pt-final}) with the appropriate Green's function below:
\begin{align}
    &G_x^{N,peri}(t) = \sum_{n=-\infty}^{\infty} G_{x+n N}(t), \\
    &\begin{aligned} \label{Gx_open}
    G_x^{N,open}(t) = \sum_{n=-\infty}^{\infty} \Big(& G_{x+2n( N+1)}(t) \\
    & - G_{-x+2n (N+1)}(t) \Big)\,.
    \end{aligned} 
\end{align}
The case of open boundary conditions is solved using the method of images by starting with a periodic ring of size $2N+2$. This ring is constructed by adding a new spin at $j=0$ and doubling the length of the spin chain giving a total of $2(N+1)$ spins. We then antisymmetrize, $r_p(t) = -r_{2N+2-p}(t)$, to enforce $r_0(t)=r_{N+1}(t)=0$ giving \C{Gx_open}.

We now turn to the equal time higher $n$-point functions, whose equation of motion is found using (\ref{Eq-t-n-pt DE}):
\begin{align} \label{n-point-eom}
    \begin{aligned}
        \dot r_{j_n\ldots j_1}
        = \sum_{p=1}^n &\Big[ -r_{j_n\ldots j_p\ldots j_1} +
        \\
        &+ \frac{\gamma_e}{2} (r_{j_n\ldots  j_{p}-1\ldots j_1}+r_{j_n\ldots  j_{p}+1\ldots j_1})
        \\
        &+ \frac{\gamma_o}{2} (r_{j_n\ldots  j_{p}-1\ldots j_1}-r_{j_n\ldots  j_{p}+1\ldots j_1}) \Big] \,.
    \end{aligned}
    \end{align}
Following the same steps as before we obtain the general solution of \C{n-point-eom} in closed form,
\begin{align} \label{n-point-sol-hom}
    r_{j_n...j_1}(t) & = \sum_{j_n'...j_1'} r_{j_n'...j_1'}(0) \prod_{p=1}^n G_{j_p-j_p'}(t) \,.
\end{align}
We see that $G_x(t)$ of (\ref{1-point-propagator}) is the basic ingredient for all equal time $n$-point functions from which all $n$-point functions can be obtained. In essence, this system satisfies a kind of Wick's theorem. 

The solution space \C{n-point-sol-hom} is very large. Physical solutions must satisfy specific conditions, which are easiest to describe for the case of the two-point function. In that case we impose the physical constraint $r_{ii}(t)=1$. Imposing this constraint means that the equations for $r_{i-1i}$, $ r_{i+1i}$, $r_{ii-1}$ and $r_{ii+1}$ now contain constant terms which we can view as forcing terms for a homogeneous differential equation. So the physical solutions split into a sum of a homogeneous solution and an inhomogeneous solution. This structure extends to higher $n$ because when two spins of the $n$-point function are set to the same point they are again replaced by $1$ giving an $(n-2)$-point function. This is highly reminiscent of the OPE (Operator Product Expansion) structure of conformal field theories.

So to find the complete solution of the problem, we also need to find the particular solution of (\ref{n-point-eom}). For the two-point function, we use the ansatz $r^{part}_{ij} = \zeta^{|i-j|}$ which is motivated by translation invariance of the equilibrium configuration. Plugged in (\ref{n-point-eom}) for $n=2$ gives $\zeta = \tanh (\beta J)$. To this solution, we can add a homogeneous solution with the property $r_{ii}^{hom}=0$:
\begin{align} \label{equal-two-point-sol}
\begin{aligned}
    r_{ij}(t) = \,&\zeta^{|i-j|} + \sum_{m,n=-\infty}^{ \infty} \Big(r_{mn}(0) - \zeta^{|m-n|} \Big)\\
    &\times \Big( G_{i-m}(t) G_{j-n}(t) - G_{i-n}(t) G_{j-m}(t) \Big)\, .
\end{aligned}
\end{align}
Note the full solution is still built out of $G_x(t)$.

\begin{figure}[h]
    \centering
    \includegraphics[width=1\linewidth]{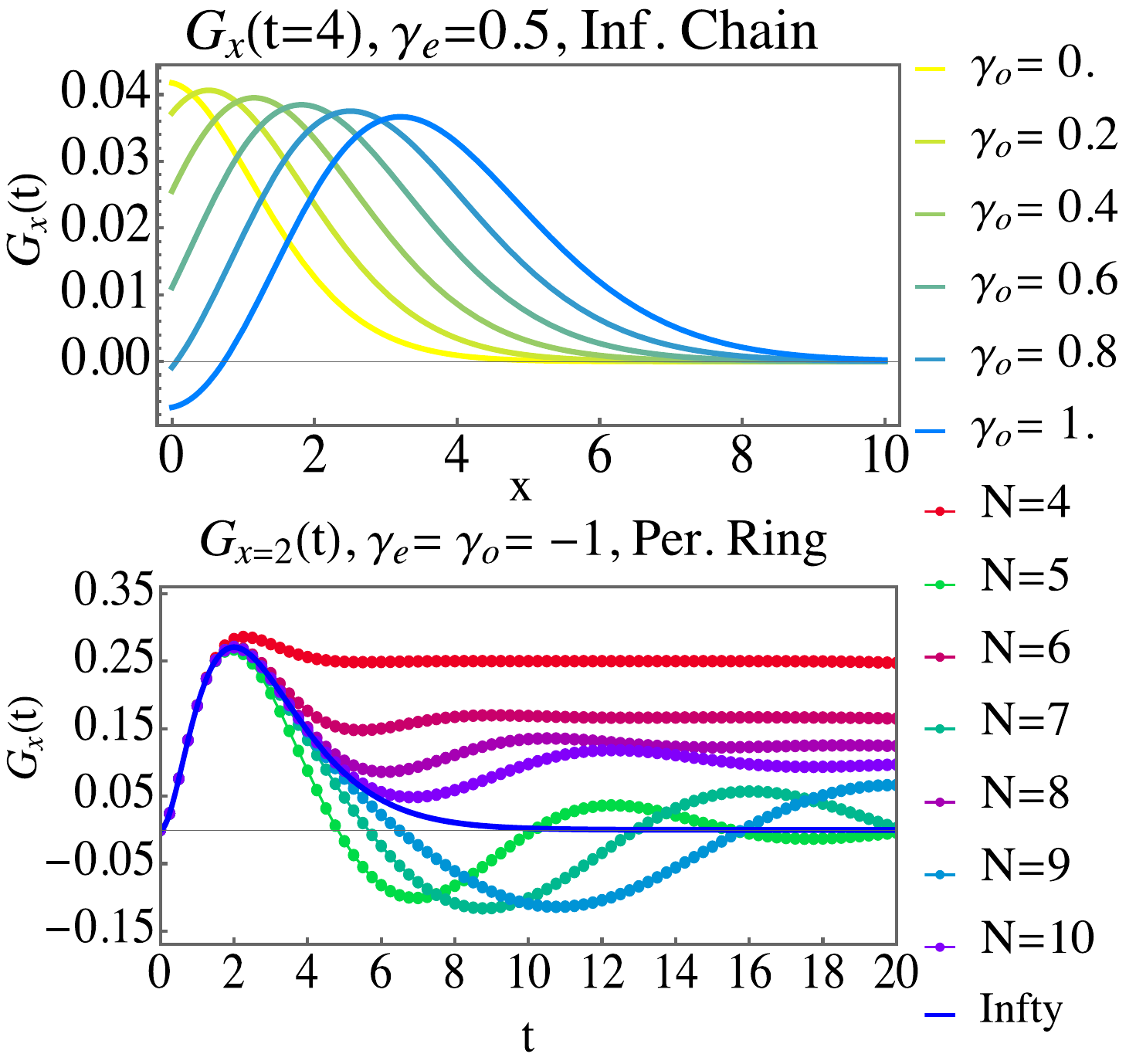}
    \caption{\textbf{Top:} Plot of $G_x(t=4)$ as a function of $x$ a for an infinite chain with $\gamma_e=0.5$ and $\gamma_o $ in the range $[0,1]$. \textbf{Bottom: } Plot of $G_{x=2}(t)$ for $\gamma_e=\gamma_o = -1$ for periodic rings of size $N$, where $N$ ranges from $4$ to $10$, compared to the infinite case.}
    \label{fig:infx_pert}
\end{figure}

The closed form expression for $G_x(t)$ allows us to study this system in complete detail at any point in the parameter space of $(\g_e, \g_o)$. In Figure \ref{fig:infx_pert} we illustrate some features of Green's function.
In the top plot with $\gamma_o \neq 0$, the excitation in the system moves to the right because of non-reciprocity. Larger $\gamma_o$ results in faster movement of the excitation. 

In the bottom plot where $\gamma_e=\gamma_o=-1$, we see that the system  displays behaviour dependent on the parity of $N$. While this behaviour is expected for anti-ferromagnetic systems of finite size, non-reciprocity modifies this behavior non-trivially so that even cases with $\gamma_e > 0$ display parity dependence for sufficiently large $\gamma_o$. With non-reciprocity, the parity dependence is noticeable when $\gamma_e - |\gamma_o| < 0$ for positive $\gamma_e$. The case $\gamma_e < 0$ appears mostly indistinguishable from this regime in terms of parity dependence.
A more detailed discussion can be found in \cite{toappear2}.

We now turn to the low energy and long wavelength behaviour of the system. This data can be extracted from the general two-point function,
\begin{align}
     C_{x}(t,\tau) = \sum_{s',s} s'_{i+x} P(s',t+\tau|s,\tau) \, s_i P(s,\tau) \,.
\end{align}
We first consider the $\tau=0$ case where $C_x(t,0)= G_x(t)$. If we let the number of sites go to infinity while the lattice spacing goes to zero, and we take the limit of very low momentum in (\ref{1-pt-momentum}), we find
\begin{align} \label{Low-E-Gx}
    G_x(t)
    & \approx e^{-(1-\gamma_e) t} \sqrt{\frac{2\pi}{ \gamma_e t}}\exp \left[-\frac{(x+\gamma_o t)^2}{2\gamma_e t} \right] \,.
\end{align}
So the relaxation time scale, correlation length and gap are given by:
\begin{align} \label{Gap-etal}
    \tau_{eq} = \frac{1}{1-\gamma_e}, && \xi_{eq} = \frac{1}{2(1-\gamma_e)}, && \mu^2 = 2(1-\gamma_e)\, .
\end{align}
Notice that $\xi_{eq}$ is the correlation length of the spacetime variable $|x+\gamma_o t|$. If $\gamma_o\rightarrow 0$ with $\gamma_e=1$, this two-point function is covariant under the scale transformation
\begin{align} \label{Scale-Transf-Age}
    x \longrightarrow \lambda^{-1} x, \qquad t \longrightarrow \lambda^{-2} t\,,
\end{align}
which is part of the Ageing group \cite{henkel_aging_2001,henkel_phenomenology_2002,henkel_non-equilibrium_2010}, but non-reciprocity violates this scale invariance. Nonetheless the system is still at criticality. Instead of \C{Scale-Transf-Age} we note that there is
still a good scale transformation for the two-point function, which now is more like a combination of a boost and scaling:
\begin{align} \label{Scale-Transf-Age-2}
    x \rightarrow \lambda^{-1}\left(x+ (1-\lambda^{-1}) \gamma_o t \right) \, , \qquad t \rightarrow \lambda^{-2} t\,.
\end{align}

While the general two-point function can be obtained by convolving (\ref{1pt-final}) and (\ref{equal-two-point-sol}), it is actually easier to directly solve the equation of motion for the two-point function,
\begin{align}
\begin{aligned}
    \frac{d}{dt}  C_{ji} = & - C_{ji} + \frac{\gamma_e}{2}(C_{j-1i} + C_{j+1i}) \\
    &+ \frac{\gamma_o}{2}(C_{j-1i} - C_{j+1i}) \,.
\end{aligned}
\end{align}
Notice that the initial condition for this differential equation is $C_{ji}(t=0,\tau) = r_{ji}(\tau)$. We can solve this equation using Fourier and Laplace transformations on the two indices $(i,j)$ and the two times $(t,\tau)$. We get
\begin{align} \label{2-pt-Integral-Form}
\begin{aligned}
     C_{kq}(s_t,s_\tau)  =& \frac{2\pi\delta_{k,-q}}{\big(s_t+1-\g_e \cos k +i \g_o \sin k\big)} \cr &\times \frac{\sqrt{(s_\tau+2)^2-4\g_e^2}  }{s_\tau\big(s_\tau+2-2\g_e \cos k\big)} \, .
\end{aligned}
\end{align}
We take the low momentum limit and inverse transform $k$, $q$ and $s_t$. We further take the $\tau\gtrsim\tau_{eq}$ limit by approximating $s_{\tau}^2 \approx0$, and a low temperature limit by approximating $2(1-\gamma_e)\approx 1-\gamma_e^2$. We then find
\begin{align} \label{Cx(t,s)-Low-T}
\begin{aligned}
    &C_{x}(t,s_\tau) = \frac{1}{s_\tau\sqrt{2\pi \gamma_e}} \\
    &\int\limits_{\frac{t}{2}(s_\tau+\mu^2)}^\infty \frac{dw}{\sqrt{w}}\exp{\left[-w- (s_\tau+\mu^2)\frac{(\gamma_o t + x)^2}{4 \gamma_e} \frac{1}{w}\right]}\, .
\end{aligned}
\end{align}
In what follows by the equilibrium limit we mean $\tau\rightarrow\infty$, which can be enforced in Laplace space by $C_x(t,eq)=\lim_{s_\tau\rightarrow 0} s_\tau C_x(t,s_\tau)$. Similarly the early time limit $t \ll \tau_{eq}$ is implemented by $t(s_\tau+\mu^2)\approx0$.
\vskip 0.1in
\noindent
\textbf{(0) \textit{Equilibrium:}} in the equilibrium limit, the two-point function is given at early times, $t\ll\tau_{eq}$ by, 
\begin{align} \label{2-pt-low-t-tau-eq}
\begin{aligned}
    C_x(t,eq) = \frac{1}{\sqrt{2\gamma_e}} \exp\left(-\mu \frac{|\gamma_ot + x|}{\sqrt{\gamma_e}} \right) \,,
\end{aligned}
\end{align}
At later times, this is replaced by a more complicated expression that can be found in \cite{toappear2}.

\vskip 0.1in
\noindent
\textbf{(I) \textit{Ageing or domain growth regime}:} we take $t \ll \tau_{eq}$ and set the temperature to zero by plugging in $\mu^2=0$. We find
\begin{align} \label{2-pt-ageing}
    C_x(t,\tau) = \frac{1}{\sqrt{2\gamma_e}} \text{erfc}\left( \frac{|\gamma_o t + x|}{2 \sqrt{\gamma_e \tau}} \right) \,.
\end{align}
If we extend (\ref{Scale-Transf-Age-2}) to include $\tau \rightarrow \lambda^{-2} \tau$ the system is scale invariant. There is also an additional scale transformation
\begin{align} \label{Scaling-Non-Trivial}
    x \rightarrow \lambda^{-1} x\,, && t \rightarrow \lambda^{-1} t\,, && \tau \rightarrow \lambda^{-2} \tau\,,
\end{align}
which also leaves (\ref{2-pt-ageing}) invariant.

\vskip 0.1in
\noindent
\textbf{(II) \textit{Approaching equilibrium}:} if we take $t\ll \tau_{eq}$ and $\tau \gg \tau_{eq}$, we see that the system approaches equilibrium as follows,

\begin{align} \label{2-pt-Approx}
\begin{aligned}
    C_x(t,\tau) = \frac{1}{\sqrt{2\gamma_e}}\bigg[&\exp\left( -\mu \frac{|\gamma_o t + x|}{\sqrt{\gamma_e}} \right) \\
    &-
    \frac{\mu |\gamma_o t + x|}{\sqrt{4\pi \gamma_e}}\frac{\exp\left(-\mu^2 \tau\right)}{(\mu^2\tau)^{3/2}} \bigg]\,.
\end{aligned}
\end{align}

\vskip 0.1in
\noindent
\textbf{(III) \textit{Spatio-temporal Porod}:} this is another regime of interest given by $|\gamma_o t + x| \ll \xi_{eq}$ \cite{bray_theory_1994}. We first take $t\ll\tau_{eq}$ and then $|\gamma_o t + x| \ll \xi_{eq}$ to get,
\begin{align} \label{2-pt-Porod}
\begin{aligned}
    C_x(t,\tau) = \frac{1}{\sqrt{2 \gamma_e}} \bigg[1 -&\bigg( \frac{e^{-\mu^2 \tau}}{\sqrt{\pi \tau}} \\
    &+\mu\ \text{erf}\left(\mu \sqrt{\tau}\right)  \bigg) \frac{|\gamma_o t + x|}{\sqrt{\gamma_e}}\bigg]\,.
\end{aligned}
\end{align}
At very low temperatures where $\mu\approx 0$, we get
\begin{align} \label{2-pt-Porod-T=0}
    C_x(t,\tau) = \frac{1}{\sqrt{2\gamma_e}} \left[ 1 - \frac{|\gamma_o t + x|}{\sqrt{\pi \gamma_e \tau}} \right]\,,
\end{align}
and the system is scale invariant under (\ref{Scale-Transf-Age-2}) extended by $\tau \rightarrow \lambda^{-2} \tau$, and by (\ref{Scaling-Non-Trivial}). If we first take the equilibrium limit $(\tau\rightarrow\infty)$ and then let the system evolve in time, the two-point function for the spatio-temporal Porod regime  is
\begin{widetext}
\begin{align} \label{2-pt-Porod-Equilibrium}
    C_x(t,eq) \approx \frac{1}{\sqrt{2\gamma_e}}\left[ 1- \left( \frac{\exp\left( - \frac{(\gamma_ot+x)^2}{2\gamma_e t} \right)}{\sqrt{\pi}\frac{|\gamma_ot+x|}{\sqrt{2\gamma_e t}}} + \text{erf} \left( \frac{|\gamma_ot+x|}{\sqrt{2\gamma_e t}} \right) \right) \mu \frac{|\gamma_o t+x|}{\sqrt{\gamma_e}} \right]\,,
\end{align}
\end{widetext}
which is covariant under (\ref{Scale-Transf-Age-2}).

These equations are very similar to what you would find if you were studying a reciprocal kinetic Ising model \cite{godreche_response_2000}. The fundamental difference is that non-reciprocity deforms the the two point functions by the boost (\ref{Boosts}) for sufficiently small times, $t\ll\tau_{eq}$. In some cases, like (\ref{2-pt-Porod-Equilibrium}), this replacement is valid for all times. We also see that the scale transformation (\ref{Scale-Transf-Age}) is generically not a symmetry of the two-point functions but we identify two regimes where the non-trivial scale invariance (\ref{Scale-Transf-Age-2}) with $\tau\rightarrow \lambda^{-2}\tau$ and (\ref{Scaling-Non-Trivial}) are symmetries. These are the ageing regime (\ref{2-pt-ageing}), and the zero temperature spatio-temporal Porod regime (\ref{2-pt-Porod-T=0}). These novel scaling properties appear to be a direct consequence of non-reciprocity in a way that has not been observed before. 
\\

\noindent
\begin{acknowledgements}
We would like to thank Michel Fruchart for helpful comments. S.~S. would like to thank the organizers of the 21st Simons Physics Summer Workshop and the ``The Landscape vs. the Swampland'' ESI workshop for hospitality during the completion of this work. M.~S. is supported in part by the Enrico Fermi Institute and in part by NSF Grant No. PHY2014195. G.~W. and S.~S. are supported in part by NSF Grant No. PHY2014195 and by the Australian Research Council (ARC) Discovery Project DP240101409.
\end{acknowledgements}
\newpage
\appendix

\section{Lightning review of the master equation}
\label{App-Master}

Even though (\ref{Markov-C}) is a good conceptual starting point for our discussion, this expression is not very well-suited for explicit calculations. Therefore we use the equivalent framework of the master equation, which is a continuous time description of (\ref{Markov-C}). We will assume at most a single spin flips at each time step. The master equation determines how the probability evolves in time, 
\begin{align} \label{Master-Eq}
   \partial_t P(s,t) &= \sum_j \Big[ w(s_j \leftarrow -s_j)P(s_1,\ldots,-s_j,\ldots,s_N;t) \nonumber  \\ & - w(-s_j \leftarrow s_j) P(s_1,\ldots,s_j,\ldots,s_N;t) \Big]\, ,
\end{align}
where $w(-s_j \leftarrow s_j)$ is the transition rate for the spin flipping from $s_j$ to $-s_j$. This equation is more general than the Markov chain (\ref{Markov-C}) because the transition rates do not need to be compatible with (\ref{Markov-C}). Indeed all that is needed for (\ref{Master-Eq}) to define a good probability theory is that
\begin{align}
    \sum_{s_j} w(-s_j \leftarrow s_j) = \alpha\,,
\end{align}
where $\alpha$ is the characteristic time in which a single spin flips; we set $\alpha=1$ in the main text. Using the master equation, we can show that the equal time $n$-point functions
\begin{align} \label{Equal-t-n-pt}
    r_{i_n ... i_1 }(t) = \langle s_{i_n} \ldots s_{i_1} \rangle (t) = \sum_s s_{i_n} ... s_{i_1} P(s,t) \,,
\end{align}
satisfy the equation of motion:
\begin{align} \label{Eq-t-n-pt DE}
    \frac{d}{dt} r_{i_n \ldots i_1 }(t)
    & = -2 \Big\langle s_{i_n} \ldots s_{i_1} \sum_{j=1}^n w(-s_{i_j}\leftarrow s_{i_j}) \Big\rangle\,. 
\end{align}
We now need  transition rates that are compatible with (\ref{Markov-C}). The transition rate for flipping spin $s_j$ assuming the system is in state $s$ at time $t-\delta t$ is,
\begin{align}
    \begin{aligned}
    w(s_j\leftarrow -s_j) & = \frac{\alpha}{2} \left( 1 + s_j \tanh h_j \right) \,  ,\\[8pt] \label{Transition-Rate-Mark}
    w(-s_j\leftarrow s_j) & = \frac{\alpha}{2} \left( 1 - s_j \tanh h_j \right)\,. 
    \end{aligned}
\end{align}
The equations of motion for the one-point function are then
\begin{align} \label{1-point-eom-h}
    \frac{d}{dt} r_j & = -\alpha r_j + \alpha \langle s_j \tanh h_j \rangle\,.
\end{align}
Since $s_j\in\{\pm1\}$ we can simplify the hyperbolic tangent,
\begin{align} \label{tanh-in-NR}
    \tanh h_j = \frac{\gamma_e}{2} \big(s_{j-1}+s_{j+1}\big) + \frac{\gamma_o}{2} \big( s_{j-1}-s_{j+1} \big)\,,
\end{align}
and then setting $\alpha=1$ gives (\ref{1-point-eom}).

\section{Useful mathematical identities} \label{Appendix-Bessel}

\subsubsection*{\ul{\it Generalization of the Sung and Hovden identity}}

We would like to prove the identity,
\begin{align} \label{Sung-Generalized}
    f_{p,N}(x,y) &= \sum_{m=0}^{N} \frac{1}{N} \exp\left[i\frac{2\pi m}{N}  p + x \cos \frac{2\pi m}{N}  - i y \sin \frac{2\pi m}{N}  \right] \,,\cr &= \sum_{n,s=-\infty}^\infty (-1)^s I_{p+s+n N}(x) J_s(y) \,,
\end{align}
which is a generalization of the Sung and Hovden identity \cite{sung_infinite_2022}. The first step is to use the Jacobi-Anger identity in appropriate places~\cite{Watson-Tretesie-1922}:
\begin{align}
\begin{aligned}
    &e^{i z \cos \theta} = \sum_{n=-\infty}^{\infty} i^n J_n(z) e^{i n \theta} \,,  \\
    &e^{i z \sin \theta} = \sum_{n=-\infty}^{\infty} J_n(z) e^{i n \theta}\,.
\end{aligned} \nonumber
\end{align}
In our specific case,  we have
\begin{align}
    \quad e^{x \cos \theta} & = \sum_{n=-\infty}^{\infty} i^n J_n(-ix) e^{i n \theta} = \sum_{n=-\infty}^{\infty} I_n(x) e^{in\theta} \,, \nonumber \\
    \quad e^{-iy \sin \theta} & = \sum_{n=-\infty}^{\infty} J_n(-y) e^{i n \theta} = \sum_{n=-\infty}^{\infty} (-1)^n J_n(y) e^{in\theta}\,,  \nonumber
\end{align}
where we have used:
\begin{align}
\begin{aligned}
    J_n(-x) & = (-1)^n J_n(x)\,,\\
    I_n(x) & = i^{-n} J_n(ix)\,, \\
    i^nJ_n(-ix) & = (-1)^n i^{2n} i^{-n} J_n(ix) = I_n(x)\,.
\end{aligned} \nonumber
\end{align}
Now we evaluate \C{Sung-Generalized} using $k=\frac{2\pi m}{Na}$ with $a$ the lattice spacing, 
\begin{align}
\begin{aligned}
    f_{p,N}(x,y) & = \sum_{k} \frac{1}{ N} e^{ika p} \sum_{r=-\infty}^\infty I_r(x) e^{i r ka } \sum_{s=-\infty}^\infty (-1)^s J_s(y) e^{i s ka}\,, \\
    & = \sum_{r,s=-\infty}^\infty (-1)^s I_r(x) J_s(y) \sum_{k} \frac{1}{ N} e^{ika(p+r+s)} \,.
\end{aligned} \nonumber
\end{align}
For $k=\frac{2\pi m}{Na}$ with $m=0,1, \ldots,N-1$ we use
\begin{align}
    \sum_{k}\frac{1}{N} e^{ika(p+r+s)} &= \sum_{m=0}^{ N-1}\frac{1}{N} e^{i\frac{2\pi m}{N}(p+r+s)} \nonumber \\
    &= \delta_{0,p+r+s+n N }, \nonumber
\end{align}
to see that the summation is non-zero only if $p+r+s= - n N$ for any integer  $n$. Therefore we get
\begin{align}
    f_{p,N}(x,y)
    & = \sum_{r,s=-\infty}^\infty (-1)^s J_s(y) I_r(x) \delta_{r,-p-s-nN} \,, \nonumber \\
    & = \sum_{n,s=-\infty}^\infty (-1)^s J_s(y) I_{-p-s-n N}(x) \,,\nonumber \\
    & = \sum_{s=-\infty}^\infty (-1)^s J_s(y) I_{p+s+n N}(x)\,, \nonumber
\end{align}
as desired. In the last step we used
\begin{align}
    I_{-n}(x) & = i^n J_{-n}(ix) = i^{2n} i^{-n}(-1)^n J_n(ix)\,, \nonumber \\
    & = i^{-n} J_n(x) = I_n(x)\,. \nonumber
\end{align}

\subsubsection*{\ul{\it Graff's generalization of Neumann's Addition Theorem}}

Graff's generalization of Neumann's Addition Theorem is given by~\cite{Watson-Tretesie-1922}
\begin{align} \label{Graff-Neumann}
    &\sum_{s=-\infty}^\infty J_{s+p}(Z) J_s(z) e^{is \phi} = J_p(\omega)  \left(\frac{Z-z e^{-i\phi}}{Z-z e^{i\phi}} \right)^{\frac{p}{2}}\,,
\end{align}
where $\omega = \sqrt{Z^2+z^2 - 2Z z \cos \phi}$. In applying this expression to (\ref{exact-1pt}) it is important to notice that this result only holds if $|Z| > |ze^{\pm i\phi}|$. Therefore the summation must be evaluated differently for the two distinct cases $|\gamma_e|>|\gamma_o|$ and $|\gamma_e|<|\gamma_o|$. To get $|\gamma_e|=|\gamma_o|$ we analytically continue the other two solutions. Performing the summation and analytic continuations carefully, we see that the one-point function can be written as (\ref{1-point-propagator}).

\newpage


\newpage

\providecommand{\href}[2]{#2}\begingroup\raggedright\endgroup

\end{document}